\documentclass[pra,twocolumn,tightenlines]{revtex4}
\usepackage{multirow}
\usepackage{bm,dcolumn,amsmath,graphicx}

\usepackage{graphicx}
\usepackage{dcolumn}
\usepackage{bm}

\begin{document}


\title{ Thallium $7p$ lifetimes derived from experimental data \\ and \emph{ ab-initio} calculations
of scalar polarizabilities }

\author{M. S. Safronova$^{1,2}$ and P. K. Majumder$^3$}
\affiliation{$^1$Physics Department, University of Delaware, Newark, DE 19716 \\
$^2$Joint Quantum Institute, NIST and the University of Maryland, Gaithersburg, MD, 20899\\
    $^3$Physics Department, Williams College, Williamstown, MA 01267}


\date{\today}

\begin{abstract}
Two different theoretical methods have been used to complete a new  calculation of polarizability in the thallium
$6p_{1/2}$, $7s$, and $7p_{1/2}$ states. The predictions of the two methods agree to within 1\% for the $6p_{1/2}$ and
$7s$ states and 2\% for $7p_{1/2}$ state. We find that
 the theoretical expression for the $6p_{1/2} - 7s$ transition polarizability difference, $\Delta \alpha_0$, is dominated
 (greater than 90\% contribution) by mixing of the $7s$ state with the $7p_{1/2}$ and $7p_{3/2}$ states.
   By comparing the theoretical expression to an existing measurement  of $\Delta \alpha_0$ [Doret \emph{et al.},
   Phys. Rev. {\bf A} 66, 052504 (2002)], new, highly-accurate values for the thallium $7p$ excited-state lifetimes
   have been extracted.  The scalar polarizability of the $7p_{1/2}$ state is also computed, anticipating an experimental
    determination of this quantity, which will then enable a high-precision determination of the $6d_{j}-7p_{j^{\prime}}$
    transition rates and provide a benchmark test of the two theoretical approaches in the near future. \end{abstract}

\maketitle


\section{\label{sec:level1} Introduction}
Thallium has played an important role in atomic-physics-based tests of discrete symmetry violation over recent decades\cite{Regan,Vetter,PorSafKoz12}.  The size of these symmetry-violating observables scales rapidly with the atomic number, encouraging the use of high-Z systems.  This therefore requires independent, precise atomic wave function calculations in order to distinguish quantum mechanical effects from the elementary particle physics observables being targeted.  For example, theoretical uncertainties in \emph{ab initio} wave function calculations in thallium currently limit the quality of the standard model test provided by a 1995 thallium parity nonconservation measurement\cite{Vetter,KozPorJoh01}.
Similarly, high-precision atomic theory is essential to interpret results from searches for atomic and molecular electric dipole moments (EDMs), as evidenced by recent calculations of the thallium EDM enhancement factor\cite{PorSafKoz12,NatSahDas11,DzuFla09,LiuKelly92}.   Independent, high-precision atomic structure measurements serve as an important tool in testing the accuracy and guiding the refinement of theoretical techniques for multi-electron systems such as thallium. Over the recent years, we have completed precise measurements of thallium transition amplitudes\cite{maj99}, hyperfine splittings\cite{maj00}, and polarizability\cite{DorFriSpe02} which show excellent agreement with theory\cite{Saf05, SafJohSaf06}.  More recently, a similar theoretical approach to that used for thallium has also been applied to other trivalent Group IIIA systems such as indium and gallium\cite{Saf07, SafSafPor13}.

Very recently, theoretical and experimental work have come together in the indium atomic system.  A measurement of the
Stark shift within the indium 410 nm $5p_{1/2} -  6s_{1/2}$ transition\cite{RanSchLor13} yielded a value for the
$6s-5p_{1/2}$ polarizability difference with 0.3\% uncertainty.  At the same time, a new \emph{ab initio} theory
effort, using two complementary, high-precision techniques, yielded a theoretical value for this quantity in excellent
agreement with the experimental result, and with 2\% estimated uncertainty\cite{SafSafPor13}.  Because the theoretical
expression for the $6s$ polarizability is dominated by terms involving the $6s-6p_{1/2}$ and $6s-6p_{3/2}$ mixing, we
show in Refs.~\cite{RanSchLor13,SafSafPor13} that a comparison of experimental and theoretical results can produce new
values for the $6p$-state lifetimes with uncertainties below 1\%.

In this paper, a similar approach of combining high-precision calculation and experiment is applied to thallium.
 An extensive calculation including uncertainties is undertaken using both a coupled-cluster (CC) as well as a
 configuration interaction + all-order (CI+all) approach to compute the polarizability of the thallium $6p_{1/2}$ ground state,
 as well as the $7s$ and $7p_{1/2}$ excited states.  We use a comparison of this theory to the 2002 thallium
Stark shift result\cite{DorFriSpe02} to extract the most precise values to date for the thallium $7p$-state lifetimes.
We also outline ongoing experimental work which will allow precise measurements of excited-state Stark shifts in both
thallium and indium.  Such measurements will then be combined with theoretical polarizability results to accurate
predict the thallium $6d-7p$ and indium $5d-6p$ transition rates.

\section{Calculation of  polarizabilities}

The valence static polarizabilities of a Tl atom can be calculated as sum over states:
\begin{eqnarray}
\alpha_0 = \frac{2}{3(2J+1)} \sum_{n} \frac{|\langle J ||D|| J_n \rangle|^2} {E_{n}-E}, \label{eq1}
\end{eqnarray}
where the sum over $n$ runs over all states with allowed $\langle J ||D|| J_n \rangle$ electric-dipole transitions.

 The $7s-6p_{1/2}$ Stark shift in Tl (i.e. the difference of the $7s$ and $6p_{1/2}$ polarizabilities)
is strongly dominated by the contributions from the $7s-7p_{j}$ transitions to the $7s$ polarizability. Therefore,
accurate measurement of this Stark shift carried out in ~\cite{DorFriSpe02} can be used to extract $7s-7p_{j}$ matrix
elements if all other smaller contributions to the $6p_{1/2}$ and $7s$ polarizabilities are calculated. The extraction
of the matrix elements also require the evaluation of relevant theoretical uncertainties.  Combining $7s-7p_j$ matrix
elements with experimental transition energies \cite{RalKraRea11} gives the $7p_{1/2}$ and $7p_{3/2}$ lifetimes.
\begin{table}
\caption{Contributions to the $7s$, $6p_{1/2}$, and $7p_{1/2}$ static polarizabilities are given in $a_0^3$ in columns
labeled ``$\alpha_0$'' . The experimental energies \cite{RalKraRea11} (in cm$^{-1})$ and the theoretical
electric-dipole reduced matrix elements (in a.u.) used to calculate dominant contributions are listed in columns
labeled ``$\Delta E$'' and ``$D$''. The CC and CI+all-order electric-dipole matrix elements and the polarizability
contributions are listed in columns labeled ``CC'' and ``CI+All'', respectively.}  \label{tab1}
\begin{ruledtabular}
\begin{tabular}{lcrrrr}
\multicolumn{1}{l}{Contribution} &  \multicolumn{1}{c}{$\Delta E$} & \multicolumn{2}{c}{$D$} &
\multicolumn{2}{c}{$\alpha_0$} \\
 \multicolumn{1}{c}{} & \multicolumn{1}{c}{Expt.}&  \multicolumn{1}{c}{CC} &\multicolumn{1}{c}{CI+All}&
  \multicolumn{1}{c}{CC} &\multicolumn{1}{c}{CI+All} \\
 \hline
          \multicolumn{6}{c}{$7s$ polarizability}  \\
          \hline
$6p_{1/2}$     & -26478&  1.826 & 1.798&-9.2(4)&-8.9\\
$7p_{1/2}$     & 7682  &  6.016 & 6.050&345(12)&349 \\
$8p_{1/2}$     & 14891 &  0.706 & 0.693&2.4(5) &2.4 \\
$(n>8)p_{1/2}$ &       &        &      &0.7(2) &    \\ [0.3pc]
$6p_{3/2}$     & -18685&  3.397 & 3.395&-45(2) &-45 \\
$7p_{3/2}$     & 8684  &  8.063 & 8.108&548(22)&554 \\
$8p_{3/2}$     & 15263 &  1.474 & 1.509&10(1)  &11  \\
$(n>8)p_{1/2}$ &       &        &      &5(1)   &    \\  [0.3pc]
Core           &       &        &      &24(1)  & 5   \\
Total          &       &        &      &881(25)& 887 \\
 \hline
\multicolumn{6}{c}{$6p_{1/2}$ polarizability}\\
\hline
 $7s           $& 26478& 1.826 & 1.798& 9.2(5)    &8.9  \\
 $8s           $& 38746& 0.535 &      & 0.54(5)   &     \\
 $(n>8)s       $&      &       &      & 0.8(3)    &     \\ [0.3pc]
 $6d_{3/2}     $& 36118& 2.334 & 2.377& 11.0(4)   &  11.5\\
 $7d_{3/2}     $& 42011& 1.101 &      & 2.1(1)    &      \\
 $(n>8)d_{3/2}$ &      &       &      & 6.4(2.8)  &      \\ [0.3pc]
 Core           &      &       &      & 24.1(1.2) & 5.0   \\
vc             &      &       &      & -4.2(9)   & -0.4  \\
 Total          &      &       &      & 50.0(3.0) &  50.7 \\
 \hline
 \multicolumn{6}{c}{$7p_{1/2}$ polarizability}\\
 \hline
  $7s           $&-7682 & 6.013 & 6.050& -344(3) & -349 \\
  $8s           $& 4586 & 6.189 &      & 611(5)  &     \\
  $(n>8)s       $&      &       &      & 22(1)   &     \\ [0.3pc]
  $6d_{3/2}     $&1958  & 10.726&10.649& 4298(24)& 4237 \\
  $7d_{3/2}     $&7852  & 4.767 &      & 212(13) &      \\
  $(n>8)d_{3/2}$ &      &       &      & 95(13)  &      \\ [0.3pc]
  Core           &      &       &      & 24(1)   & 5   \\
  Total          &      &       &      & 4918(30)& 4831 \\
  Final          &     &         &      & 4918(120)&  \\
 \end{tabular}
\end{ruledtabular}
\end{table}
\begin{table*}
\caption{Final values of the $7s$ and $6p_{1/2}$ polarizabilities and their difference $\Delta \alpha_0$ (a.u.).
Determination of the reduced electric-dipole $7s-7p_j$ matrix elements (in a.u.) and $7p_j$ lifetimes (in ns) from the
combination of  measured Stark shift ~\cite{DorFriSpe02} and theoretical values. The quantity $C$ is the value of
$\Delta \alpha_0(7s-6p_{1/2})$ with the contribution of the $7s-7p_j$ transitions subtracted out.
$^a$Ref.~\cite{Mil02}, $^{b}$Ref.~\cite{DorFriSpe02}, $^c$Ref.~\cite{KozPorJoh01}, $^d$Ref.~\cite{DzuFla09},
$^e$Ref.~\cite{BorZelEli12}.} \label{tab2}
\begin{ruledtabular}
\begin{tabular}{lcccccccc}
\multicolumn{1}{l}{} &  \multicolumn{1}{c}{$\alpha_0(7s)$} & \multicolumn{1}{c}{$\alpha_0(6p_{1/2})$} &
\multicolumn{1}{c}{$\Delta \alpha_0(7s-6p_{1/2})$} & \multicolumn{1}{c}{$C$} & \multicolumn{1}{c}{$D(7s-7p_{1/2})$}&
\multicolumn{1}{c}{$D(7s-7p_{3/2})$} &\multicolumn{1}{c}{$\tau(7p_{1/2})$}&
\multicolumn{1}{c}{$\tau(7p_{3/2})$}  \\
\hline
CC                 &881     &50.0     & 831        & -61.7& 6.016    & 8.063    &  60.15        & 46.38         \\
CI+All             &887     &50.7     & 836        & -66.4& 6.050    & 8.108    &  59.48        &  45.86         \\
Final              &881(9) &50.0(1.0)& 831(8)    & -61.7(6.7)& 6.013(27)& 8.058(37)& 60.21(55)& 46.44(42)\\
Expt.      &         &      51(7)$^a$  & 829.7(3.1)$^b$&      &          &          &          &           \\
Theory     &         & 49.2$^c$,48.8$^d$ &              &      &          &          &            &         \\
Theory     &         & 52.1(1.6)$^e$ &              &      &          &          &            &         \\
\end{tabular}
\end{ruledtabular}
\end{table*}

The polarizabilities of the ground $6p_{1/2}$ and excited $7s$ states were calculated using the linearized coupled-
cluster method in \cite{SafJohSaf06}, but their uncertainties were not evaluated. In this work, we carry out additional
linearized coupled-cluster calculations (CC) to estimate the uncertainties of each term. We also carry out another
independent calculation of these polarizabilities using a recently developed hybrid approach that combines
configuration interaction and linearized coupled-cluster method (CI+all-order) ~\cite{SafKozJoh09}. This calculation
treats Tl as a system with three valence electrons and accurately accounts for configuration mixing and valence-valence
correlations. The CC calculation treats Tl as a monovalent system with $6s^2$ considered to be a part of the closed
core; however, the CC approach includes some additional high-order corrections to the dipole operator. Since these two
methods differ in their inclusion of higher-order effects, comparing their results provides additional evaluation of
the uncertainty of our calculations. We refer the reader to Refs.~\cite{SafJoh08,SafJohSaf06,SafSaf11} and
\cite{SafKozJoh09,SafPorCla12,SafKozCla11,SafKozSaf12,SafPorKoz12,PorSafKoz12} for detailed descriptions of the
linearized coupled-cluster and CI+all-order methods, respectively. The results of both methods were recently compared
for In polarizabilities in \cite{SafSafPor13}.

The breakdown of the contributions to the $7s$ and $6p_{1/2}$ Tl polarizabilities is listed in Table~\ref{tab1}.
Experimental energies are given for all terms that are listed separately, such as $7p$ contributions to the $7s$
polarizability.  The uncertainties of the main CC terms are determined from the spread  of four different
coupled-cluster calculations carried out in this work (with and without the perturbative triple terms and with
inclusion of the scaling to account for some missing higher-order corrections); CC matrix elements from
\cite{SafJohSaf06} are kept as final values. The determination of the uncertainties is described in detail in
Refs.~\cite{SafSaf11,SafSafPor13}. The ($n>8$) contributions for the $7s$ polarizability contain $9p$ contribution
calculated using  the all-order method and all other ($n>9$) terms calculated in the random-phase approximation (RPA)
and scaled to account for higher-order corrections. The scaling factor is determined as the ratio of the total CC value
for the main $n=6-9$ terms and corresponding RPA result. For the $6p_{1/2}$ polarizability, $(7-10)s$  and
$(6-9)d_{3/2}$ contributions are calculated by combining CC matrix elements and experimental energies, and the
remaining contributions are calculated together using the scaled RPA approach. The difference of the \textit{ab initio}
RPA and scaled RPA values is taken to be uncertainty of these high$-n$ contributions.
 The
ionic core polarizabilities and small (vc) term that accounts for the occupied valence shell(s) are listed separately
in rows ``core'' and ``vc''. The vc term is negligible for the $7s$ and $7p_{1/2}$ polarizabilities. Core and vc
contributions are calculated in the RPA. The differences of the Dirac-Fock and RPA values are taken to be  their
uncertainties. We note that core polarizability is much larger in the CC method, since $6s^2$ shell is included in the
core in the CC calculation, while the $6s$ shell belongs to the valence space in the trivalent CI+all-order
calculation.

 The sum-over-states is not used in the CI+all-order calculation of the
polarizabilities which is carried out by  solving the inhomogeneous equation of perturbation theory in the valence
space \cite{PorRakKoz99a}. However, we evaluated a few dominant terms separately by combining CI+all-order matrix
elements with experimental energies to compare these terms in both approaches. These results are listed in the last
column of Table~\ref{tab1}.

 We also calculated the thallium $7p_{1/2}$
polarizability using both CC and CI+all-order methods. Since the $7p_{1/2}-6d_{3/2}$ matrix element strongly dominates
$7p_{1/2}$ polarizability, our calculation can be used to extract this matrix element if either the $7p_{1/2}-6p_{1/2}$
or $7p_{1/2}-7s$ Stark shift is measured with high-precision. We have used the $7s-7p_{1/2}$ matrix element determined
in the next section to provide more accurate recommended value. We determine the contribution of all other terms except
the $7p_{1/2}-6d_{3/2}$ term to be 620(36) a.u. (see Table~\ref{tab1}).  The determination of the final uncertainties
is described in the next section.

\section{Determination of thallium $7p$ lifetimes}
\begin{figure}[t]
\includegraphics[scale = 0.4]{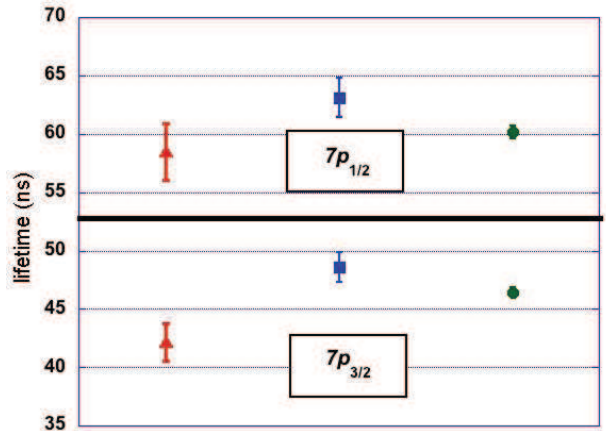}
\caption{(Color online) A comparison of the present $7p$-state lifetime determinations to older experimental results.
$7p_{1/2}$ results are shown in the top portion of the figure, and $7p_{3/2}$ results below.  References as follows:
(red) triangles: Ref. \cite{LHunter82};  (blue) squares:  Ref. \cite{James86}; (green) circles:  present work.}
\label{tl7pexp}
\end{figure}

Separating the $7s-7p_j$ contributions (see Eq.(\ref{eq1})), we write the $\Delta \alpha_0(7s-6p_{1/2})$ Stark shift as
\begin{equation}
\label{eq14} \Delta \alpha_0(7s-6p_{1/2})=B S+C,
\end{equation}
where
\begin{equation}
B=\frac{1}{3}\left( \frac{1}{E(7p_{1/2})-E(7s)} + \frac{R^2}{E(7p_{3/2})-E(7s)} \right), \label{B}
\end{equation}
$S=D^2$ is the $7s-7p_{1/2}$ line strength, $R$ is the ratio of the $D(7s-7p_{3/2})$ and $D(7s-7p_{1/2})$ reduced E1
matrix elements $R=1.340(4)$, and  term  $C$ contains all other contributions to the Stark shift.  We use the  $\Delta
\alpha_0(7s-6p_{1/2})$ measured in Ref.~\cite{DorFriSpe02}. Combining experimental energies from ~\cite{RalKraRea11}
and our theoretical value of the ratio gives $B=24.65(9)$~a.u. The final results of our calculations for $6p_{1/2}$,
$7s$ polarizabilities, their difference, and term $C$ are given in Table~\ref{tab2}. Our theoretical value for the
$7s-6p_{1/2}$ Stark shift is in excellent agreement with the experiment \cite{DorFriSpe02}. The ground state
polarizability is compared with other theory \cite{KozPorJoh01,DzuFla09,BorZelEli12} and experiment~\cite{Mil02}. Since
CC and CI+all-order values include all dominant correlation corrections between these two calculations, we estimate the
uncertainty in the dominant contributions as the difference between the CC and CI+all-order values $\delta\alpha$.
Then, we assume that all other uncertainties do not exceed  the uncertainty of the dominant corrections $\delta\alpha$.
Adding these uncertainties in quadrature, we arrive at he total uncertainty of $\sqrt{2}\delta\alpha$. The uncertainty
in the term $C$ is determined by the same procedure. The uncertainty in the $7s-7p_{1/2}$ line strength $S$ is
determined as
\begin{equation} \delta S=\frac{1}{B}\sqrt{(\delta C)^2+(\delta
\Delta\alpha_0)^2+(S \delta B)^2}.
\end{equation}
 The
lifetimes of the $7p_{j}$ states are obtained using  $\tau_a=1/A_{ab}$ , where the transition rate $A_{ab}$ is given by
\begin{equation}
A_{ab}=\frac{2.02613\times10^{18}}{\lambda_{ab}^3}\frac{S_{ab}}{2J_a+1}\,\,\text{s}^{-1};
\end{equation}
the transition wavelength $\lambda_{ab}$ is in \AA~. The final recommended values for the $7p_{1/2}$ and $7p_{3/2}$
lifetimes are listed in Table~\ref{tab2}. The purely theoretical lifetime results of the CC calculation are in very good agreement with these recommended
values.  Fig. \ref{tl7pexp} compares the present results to earlier experimental determinations of these lifetimes.

\section{Measurement of excited-state Stark shifts in thallium and indium}
 As described here, as well as in \cite{SafSafPor13}, calculations of the excited $p$-state polarizabilities,
 coupled with future measurements of excited-state Stark shifts, will allow important new tests of the atomic theory.
In particular, a  measurement of the Stark shifts involving 7p states in Tl and 6p state in In will allow a
definitive test of the CC and CI+all-order theoretical methods. While both approaches give results in very close agreement for the ground state and the first excited
 $ns$ state polarizabilities of In and Tl, the differences increase for the next excited $np$ state polarizability (to 2\% for Tl and 4\% for
 In~\cite{SafSafPor13}).
Precise Stark shift measurements involving these excited $np$ states will directly address the question of whether accurate treatment of the configuration mixing or
 higher-order corrections to the matrix elements are more important for such states. Moreover, measurement of these Stark shifts 
  will allow determination of the $7p-6d$ transition rates in Tl and $6p-5d$ transition rates in In.
 These are very important for improved theoretical descriptions of the $d$-state properties.  These rates cannot be determined accurately from
 $nd$-state lifetime measurements such as reported in \cite{taylor08} owing to very small branching ratios.  
 
 We are currently undertaking atomic beam-based experiments to measure the excited $p$-state polarizabilities in both  thallium and indium. To achieve this, having completed Stark shift experiments in the ground state 410 nm and 378 nm transitions of these two Group IIIA elements, we are now introducing a two-step, two-color spectroscopy measurement scheme for use with our atomic beam apparatus.  For both atomic systems, we have completed such two-step spectroscopy experiments in a vapor cell environment to study excited-state hyperfine structure and isotope shifts\cite{gunawardena09,rankealmaj13}. In these two-step experiments, we begin by locking the blue or UV laser to the first transition step, using a supplementary atomic vapor cell and technique developed recently\cite{GunHess08}.  We then overlap this laser beam with that of a second infrared laser and intersect both with our atomic beam using a transverse geometry.  Interestingly, the thallium $7s - 7p_{1/2}$, and the indium $6s - 6p_{1/2}$ and $6s - 6p_{3/2}$  transitions, with resonance wavelengths of 1301 nm, 1343 nm, and 1291 nm respectively, can all be reached by a single external cavity diode laser which is currently in use (Sacher Lasertechnik, model TEC-150-1300-050). Using the FM spectroscopy technique described in \cite{RanSchLor13}, will will extract high resolution spectra from the atomic beam transmission signal of the infrared laser, expected to be an order of magnitude weaker than the signal from the analogous single-step experiment,  given that here we promote only a fraction of the ground state atoms to the intermediate state.
\begin{figure}[t]
\includegraphics[scale = 0.4]{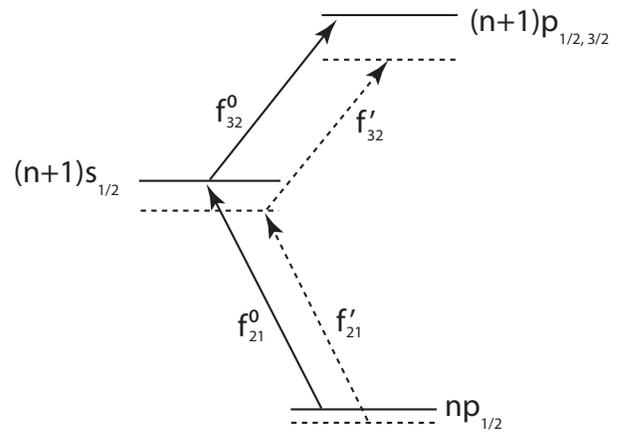}
\caption{A schematic diagram showing the configuration of low-lying energy levels of both thallium and indium.  The dotted lines suggest the Stark-shifted levels, where the size of the level shifts (not to scale) become successively larger for the higher lying states.  }
\label{tlenglevel}
\end{figure}

In general, extracting Stark shift information in a two-step transition experiment is complicated by the fact that both transitions are shifted when the static field is turned on.  Fig. \ref{tlenglevel} indicates the general form of such a three-level, two-step process in either indium or thallium.  As indicated, in the presence of an electric field, all energy levels are Stark-shifted downwards with the magnitude of the shift increasing for higher-lying states.  We define the magnitude of the Stark shift within the lower (upper) transition as $\Delta_{21}$ ($\Delta_{32}$), and the field-free resonance frequencies of these transitions as $f_{21}^0$ and $f_{32}^0$ respectively.  We note that for both elements $\Delta_{32} \gg \Delta_{21}$.   By keeping the first-step laser locked to the atomic transition in a \emph{field-free} region, the first-step excitation in the electric field region would then be shifted slightly out of resonance.  This results in the excitation of  a non-zero velocity class of atoms.  In this case, the resonance frequency for the second step transition, $f_{32}^\prime$, in the presence of the electric field is given by:
\begin{equation}
\label{eqss}
f_{32}^\prime = f_{32}^0 - |\Delta_{32}| + |\Delta_{21}| \frac{f_{32}}{f_{21}},
\end{equation}
where the final term results from the Doppler shift produced by the off-resonant first-step excitation.  For electric field of 10 kV/cm, typical for these experiments, $\Delta_{21} \approx 10$ MHz, still much less than the $\sim$100 MHz residual Doppler width in the atomic beam geometry.  Thus, the decrease in excitation efficiency for the first step transition in the presence of the electric field will not be significant.  Furthermore, since the optical resonance frequencies $f_{21}$ and $f_{32}$ in Eq. \ref{eqss} are known, and $\Delta_{21}$ has been previously measured to high accuracy\cite{DorFriSpe02}, we will be able to determine $\Delta_{32}$ unambiguously. For transitions involving $J=1/2$ states, there is only a scalar component to the polarizability, but for the indium 1291 nm $6s - 6p_{3/2}$ transition, there will exist both scalar and tensor contributions, both of which were computed in \cite{SafSafPor13}.  In this case we will study each resolved hyperfine transition and will vary the laser polarization relative to the static field direction to extract both polarizability components.

\section{Conclusions}
In conclusion, through a comparison of an existing thallium Stark shift measurement and new \emph{ab initio} calculations of scalar polarizabilities, we have derived new, highly-accurate values for the thallium $7p$ excited-state lifetimes.   Recent measurements of Stark shifts in thallium and indium serve an important function as benchmark tests of two distinct atomic theory techniques that can be applied to these multivalence systems.  Future measurements of excited-state Stark shifts in both elements will test the atomic theory approaches in important new ways since the relevant polarizability is dominated by mixing with excited $d$-states in these systems, whose theoretical contributions are more uncertain.  A theory-experiment comparison will thus allow precise derivation of thallium $7p-6d$ and indium $6p-5d$ transition matrix elements.

\begin{acknowledgments}
 We thank U.~I. Safronova and S.~G. Porsev for useful discussions. The work of M.S.S was supported in part by the NSF Grant No. PHY-1068699.  We thank  G. Ranjit, and N.A. Schine for important experimental contributions. The experimental polarizability work was supported by NSF Grant No. PHY-0140189 and is currently supported by Grant No. PHY-0969781.
\end{acknowledgments}


\bibliography{tl}

\end{document}